\documentclass[conference]{IEEEtran}
\IEEEoverridecommandlockouts
\usepackage{cite}
\usepackage{amsmath,amssymb,amsfonts}
\usepackage{algorithmic}
\usepackage{graphicx}
\usepackage{textcomp}
\usepackage{xcolor}
\usepackage{eso-pic} 
\def\BibTeX{{\rm B\kern-.05em{\sc i\kern-.025em b}\kern-.08em
    T\kern-.1667em\lower.7ex\hbox{E}\kern-.125emX}}
\begin{document}
\AddToShipoutPictureFG*{%
  \AtPageLowerLeft{%
    \put(\LenToUnit{0.5\paperwidth},\LenToUnit{0.32in}){%
      \makebox[0pt][c]{\parbox{\textwidth}{\centering\scriptsize
        \textcopyright~2026 IEEE. Personal use of this material is permitted. Permission from IEEE must be obtained for all other uses, in any current or future media, including reprinting/republishing this material for advertising or promotional purposes, creating new collective works, for resale or redistribution to servers or lists, or reuse of any copyrighted component of this work in other works.\\
        Accepted to the 2026 IEEE Frontiers in Education Conference (FIE). DOI: TBD}}}}}
\title{Virtual Global Collaboration in Data Analytics and Machine Learning Education:
A Mixed-Methods Study of Asynchronous Cross-Border Teamwork}
\author{
\IEEEauthorblockN{Sehrish Basir Nizamani\IEEEauthorrefmark{1}, Saad Nizamani\IEEEauthorrefmark{1}, Khyati Goyal\IEEEauthorrefmark{1}, Sarwat Nizamani\IEEEauthorrefmark{2}, Zannah Ziew\IEEEauthorrefmark{1}}
\IEEEauthorblockA{\IEEEauthorrefmark{1}Department of Computer Science, Virginia Tech, Blacksburg, VA, USA\\
sehrishbasir@vt.edu, saadnizamani@vt.edu, khyati22@vt.edu, zannah24@vt.edu}
\IEEEauthorblockA{\IEEEauthorrefmark{2}Department of Computer Science, University of Mirpurkhas, Mirpurkhas, Pakistan\\
sarwat@umpk.edu.pk}
}
\maketitle
\begin{abstract}
This innovative practice full paper examines a structured Virtual Global Collaboration (VGC) activity between undergraduate computing courses in the United States and Pakistan. The project was designed to support technical and intercultural skill development through asynchronous international teamwork in data analytics and introductory machine learning. Students worked in mixed-institution teams through a six-phase project including cultural orientation, dataset selection, data cleaning, analysis, introductory machine learning, and structured reporting. Using shared computational tools, teams coordinated across time zones to complete a joint data-driven project. Survey and qualitative reflection data were analyzed to examine collaborative experience, communication and cultural dynamics, learning outcomes, perceived value, and global readiness. Results show consistently positive student experiences across both cohorts, with collaborative processes strongly associated with learning and perceived value, and cross-cultural communication emerging as the primary driver of global readiness. These findings demonstrate how short-term, structured VGC can be effectively integrated into computing courses to support both technical learning and global competence.
\end{abstract}
\section{Introduction}
Virtual global collaboration (VGC) has emerged as an effective approach for integrating intercultural and international learning into higher education. Through course-embedded activities, VGC connects students with peers around the world using online platforms, enabling collaborative learning without the barriers of traditional study abroad programs.
By engaging in joint projects and structured interactions, students gain exposure to different cultural viewpoints while applying discipline-specific knowledge in authentic, real-world settings. When deliberately designed with structured reflection and explicit communication scaffolding, VGC can support intercultural competence, teamwork, and the negotiation and problem-solving skills needed to navigate cross-cultural challenges in distributed environments \cite{alvarez2019virtual}.
Building on these approaches, this study investigates a course-embedded VGC project conducted between 3000-level undergraduate computer science courses in the United States and Pakistan, respectively, within a data analytics and machine learning context. It examines how students experience and learn through primarily asynchronous, cross-border teamwork, and how these experiences influence their perceptions of collaboration, learning, and global readiness.
\section{Related Work}
VGC often implemented through Collaborative Online International Learning (COIL), has emerged as an effective approach for integrating global learning into undergraduate education. Prior research shows that virtual exchange enables meaningful cross-border collaboration while overcoming barriers of traditional study abroad \cite{vahed2020, commander2022}.
\subsection{Learning Outcomes in Virtual Global Collaboration}
Studies consistently show that virtual collaboration enhances both discipline-specific learning and transferable skills. Students participating in COIL-based experiences report improved understanding of course concepts \cite{vahed2020, woodley2023}. In addition to technical learning, these experiences support the development of critical professional competencies, including communication, collaboration, and problem-solving skills, which are essential in globally distributed professional environments \cite{dinca2023, zhang2025}.
Empirical research further suggests that virtual collaboration can lead to measurable improvements in student outcomes. For example, one quasi-experimental study found that students engaged in COIL demonstrated significantly higher intercultural competence gains than control-group students at one of two study sites and on one of two validated measures used, with no significant difference observed at the second site or on the second measure \cite{hackett2023}. These findings indicate that virtual collaboration can function as a meaningful pedagogical approach rather than simply a supplementary activity.
\subsection{Intercultural Competence and Global Awareness}
A central contribution of the VGC is its ability to enhance intercultural competence and global awareness. Students frequently report increased openness to diverse perspectives, improved cultural sensitivity, and greater awareness of global issues \cite{wood2022, simoes2023}. These experiences encourage students to reflect on their own cultural assumptions while engaging with peers from different backgrounds.
Research using validated intercultural competence measures demonstrates significant gains in students’ intercultural development following virtual exchange participation \cite{hackett2023, chu2024}. Some studies indicate that virtual exchange has downstream effects beyond the course itself; one study found that students who participated in virtual exchange showed increased subsequent participation in physical study abroad programmes \cite{case2022}. Virtual collaboration also supports the concept of “internationalization at home,” enabling broader access to global learning experiences \cite{garcia2023, simoes2023}.
\subsection{Design Considerations and Challenges}
Despite its benefits, global collaboration presents several challenges. Common issues include time zone differences, communication barriers, uneven participation, and technological limitations \cite{mlodzianowska2024, davidsonshivers2023}. These challenges can affect team coordination and overall collaboration quality, particularly in asynchronous environments.
Effective instructional design has been shown to mitigate these challenges. Structured tasks, clear expectations, and guided collaboration frameworks are critical for supporting student engagement and learning outcomes \cite{ismailov2021, rakhma2025}. While both synchronous and asynchronous collaboration can support effective learning outcomes, the choice of modality may influence student engagement and satisfaction \cite{davidsonshivers2023}.
\subsection{Summary and Research Gap}
Overall, existing literature demonstrates that VGC can enhance student learning, intercultural competence, and global awareness across disciplines. However, there is limited research examining short-term, structured virtual collaborations in computing education that integrate both data analytics and machine learning tasks, particularly in predominantly asynchronous settings.
\section{Conceptual Framework}
\begin{figure}[ht]
\centering
\includegraphics[width=0.75\linewidth]{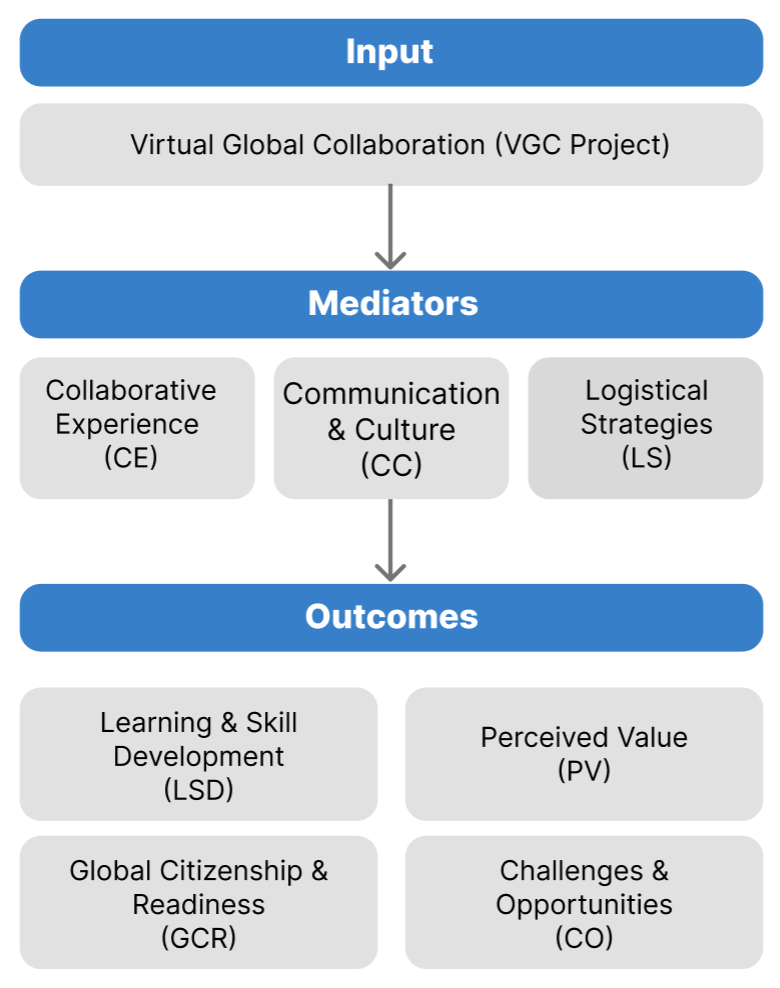}
\caption{Conceptual framework of the VGC study.}
\label{fig:conceptual_model}
\end{figure}

This study is guided by the conceptual framework shown in Fig.~\ref{fig:conceptual_model}, which shapes student experiences and learning outcomes. The framework is informed by prior research on collaborative learning, computer-supported collaborative learning, and virtual exchange, which suggests that student outcomes in technology-mediated environments depend not only on the task itself but also on the quality of interaction, communication, and coordination among team members \cite{johnson1999, vahed2020}.
In this study, VGC participation serves as the instructional context in which students work in mixed-institution teams on data analytics and machine learning tasks. The framework assumes that this experience influences student outcomes through three interconnected mediating dimensions: collaborative experience, communication and cultural dynamics, and logistical strategies. These mediators reflect key processes identified in prior work on virtual collaboration and international teamwork, where positive interdependence, cross-cultural interaction, and effective coordination are central to successful collaboration \cite{johnson1999, hackett2023, davidsonshivers2023}.
Collaborative experience refers to students’ perceptions of inclusion, participation, and overall teamwork quality. Research on collaborative learning shows that when students perceive their team experience positively, they are more likely to remain engaged and report stronger learning outcomes \cite{johnson1999}. Communication and cultural dynamics capture how students interact across cultural boundaries, including their ability to share ideas, adapt communication styles, and benefit from diverse perspectives. Prior studies of virtual exchange suggest that these interactions contribute to intercultural competence and global awareness \cite{hackett2023, chu2024}. Logistical strategies refer to how students manage practical aspects of distributed teamwork, such as time zone coordination, task allocation, and the use of shared digital tools. Research on virtual teams emphasizes that coordination and communication mechanisms are critical for sustaining effective collaboration in asynchronous settings \cite{davidsonshivers2023, mlodzianowska2024}.
These mediating processes are expected to influence three primary outcome dimensions. The first dimension, learning and skill development, refers to students’ perceived gains in technical knowledge, analytical ability, and transferable collaboration skills. The second dimension, perceived value, captures students’ overall assessment of the usefulness and relevance of the activity. The third dimension, global citizenship and readiness, refers to students’ confidence and preparedness for working in diverse, international, and distributed environments. Prior research suggests that collaborative and project-based virtual exchange can strengthen all three of these outcomes by combining authentic disciplinary work with intercultural engagement\cite{vahed2020, dinca2023, garcia2023}.
The framework also includes challenges and opportunities as a reflective outcome, reflecting whether students view difficulties as part of meaningful learning. Prior work suggests that good instructional design can help prevent common collaboration problems, such as free-rider, sucker, and silo effects, from undermining outcomes \cite{davidsonshivers2023}.
Based on this framework, the following hypotheses were proposed.

\begin{itemize}
\item \textbf{H1 Collaborative Experience:} Students participating in the VGC activity will report positive collaborative experiences.
\item \textbf{H2 Communication and Cultural Dynamics:} Students participating in the VGC activity will report effective cross-cultural communication and culturally enriching collaboration.
\item \textbf{H3 Logistical Strategies:} Students participating in the VGC activity will report that logistical challenges were manageable.
\item \textbf{H4 Learning and Skill Development:} Collaborative experience, communication and cultural dynamics, and logistical strategies will positively predict students’ perceived learning and skill development.
\item \textbf{H5 Perceived Value:} Collaborative experience, communication and cultural dynamics, and logistical strategies will positively predict students’ perceived value of the collaboration.
\item \textbf{H6 Global Citizenship and Readiness:} Collaborative experience, communication and cultural dynamics, and logistical strategies will positively predict students’ global citizenship and readiness for international collaboration.
\item \textbf{H7 Challenges and Opportunities:} Students will perceive the opportunities and growth gained from the collaboration to outweigh the challenges encountered.
\end{itemize}

\section{Study Context}
The study was conducted within undergraduate computing courses at a large public research university in the United States in collaboration with a partner public university in Pakistan. The VGC activity was embedded as an optional component within courses focused on data analytics and introductory machine learning.
\subsection{Participants}
Participants included undergraduate students from both institutions who voluntarily opted into the VGC activity. Students were organized into mixed-institution teams, ensuring that each group included members from both universities. This structure was designed to promote cross-cultural interaction, shared problem-solving, and exposure to diverse perspectives.
\subsection{Instructional Design of the VGC Activity}
\begin{figure}[ht]
\centering
\includegraphics[width=0.9\linewidth,height=6cm,keepaspectratio]{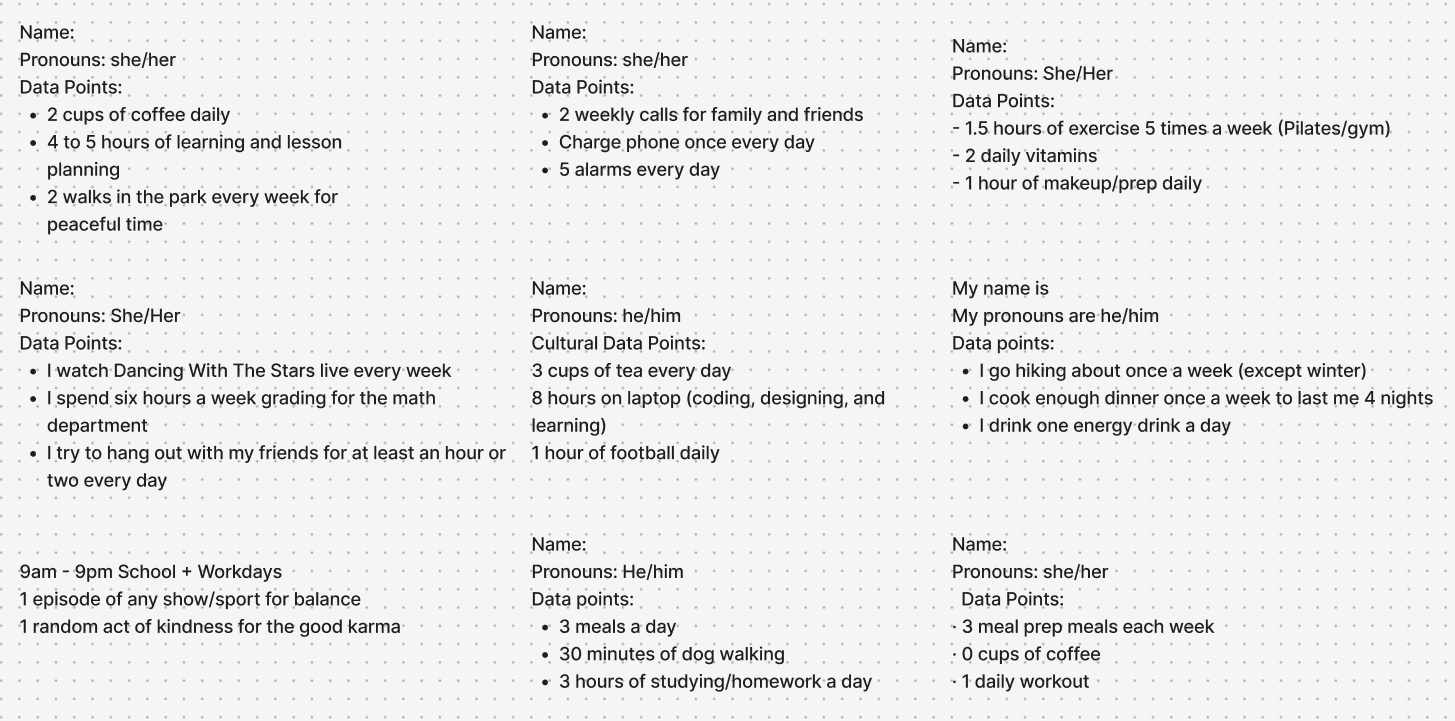}
\caption{FigJam board illustrating students’ cultural data points shared during the cultural introduction and orientation phase.}
\label{fig:figjam_intro}
\end{figure}

The VGC activity was designed as a structured, project-based learning experience integrating both technical and intercultural components. The design emphasized explicit task structuring, role clarity, and scaffolded collaboration to support communication across cultural and institutional boundaries.
The activity followed a six-phase structure:

\textbf{Phase 1: Cultural Introduction and Orientation}: Students participated in a cultural icebreaker activity using “Cultural Data Points,” where they shared short quantitative descriptions of their daily lives. This activity was complemented by short videos introducing cultural contexts, encouraging students to reflect on how culture influences communication and teamwork (see Fig.~\ref{fig:figjam_intro}).

\textbf{Phase 2: Dataset Selection and Setup}: Teams selected a dataset from an open data repository and established a shared Google Colab workspace. Students coordinated access, defined file formats, and created an initial plan for data preparation.

\textbf{Phase 3: Data Cleaning and Planning}: Teams collaboratively cleaned the dataset and documented preprocessing steps. Students developed multiple analysis ideas and divided responsibilities across institutions, with one group focusing on visualization and the other on machine learning tasks.

\textbf{Phase 4: Data Analysis and Modeling}: Students implemented their assigned tasks, including data visualization and basic machine learning modeling. Teams compared results and collaboratively interpreted findings.

\textbf{Phase 5: Structured Reporting}: Teams organized their work using a Question--Analysis--Conclusion (QAC) structure to clearly communicate their analytical process and results. A collaborative reflection component was also included.

\begin{figure}[ht]
\centering
\includegraphics[width=0.9\linewidth,height=7cm,keepaspectratio]{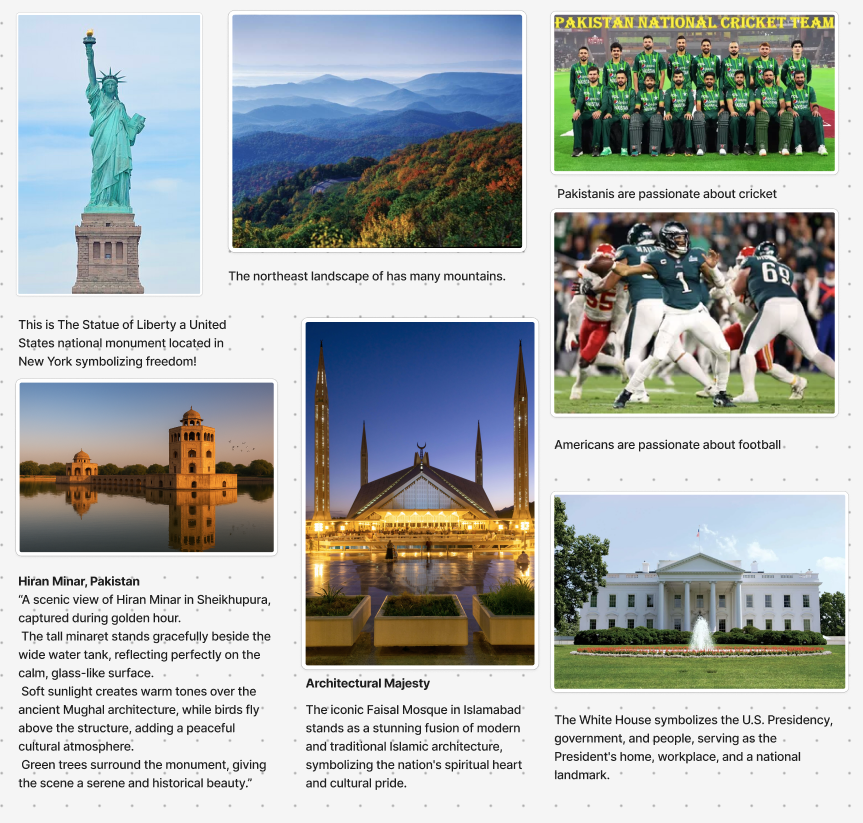}
\caption{Cultural symbols, landmarks, and sports shared by students during the reflection and closure phase}
\label{fig:culture}
\end{figure}

\textbf{Phase 6: Reflection and Closure}: Students completed and submitted a post-project reflection survey and shared brief cultural reflections on a shared Figma Jam Board, using visualizations to represent their cultural backgrounds. A snapshot is shown in Fig.~\ref{fig:culture}. Final project artifacts were submitted for review.

Students collaborated using Microsoft Teams for communication and Google Colab for coding and analysis. The design emphasized asynchronous interaction, while allowing optional synchronous meetings to support coordination when needed.
\subsection{Survey Instrument}
To evaluate student experiences, a post-project survey was administered. The survey consisted of Likert-scale items organized into seven constructs derived from the conceptual framework:
\begin{itemize}
\item Collaborative Experience (CE)
\item Communication and Cultural Dynamics (CC)
\item Logistical Strategies (LS)
\item Learning and Skill Development (LSD)
\item Perceived Value (PV)
\item Global Citizenship and Readiness (GCR)
\item Challenges and Opportunities (CO)
\end{itemize}
Responses were measured using a five-point Likert scale ranging from Strongly Disagree (1) to Strongly Agree (5), capturing both collaboration processes and learning outcomes. The complete survey instrument, including all Likert-scale
and open-ended items organized by construct, is provided
in Appendix~\ref{tab:survey}.
\subsection{Data Collection}
Survey responses were collected electronically using Microsoft Forms at the conclusion of the activity. Participation was voluntary and anonymized, and informed consent was obtained prior to participation.
\subsection{Data Analysis}
Likert responses were coded from 1 to 5, and construct scores were computed as averages of their respective items. Reliability was assessed using Cronbach’s alpha \cite{cronbach1951}.
Normality was evaluated using the Shapiro--Wilk test \cite{shapiro1965}, with all constructs showing non-normal distributions ($p < .05$); therefore, non-parametric tests were used. Descriptive statistics (means, standard deviations, medians) were computed for all constructs.
One-sample Wilcoxon signed-rank tests \cite{wilcoxon1945} assessed whether construct medians exceeded the neutral midpoint (H1--H3). A paired Wilcoxon test compared challenge-focused and opportunity-focused items (H7). Group differences between U.S. ($n = 21$) and Pakistan ($n = 15$) cohorts were tested using Mann--Whitney U tests \cite{mann1947} with rank-biserial effect sizes.
Spearman rank correlations \cite{spearman1904} were computed among all seven construct means to examine bivariate relationships between mediator and outcome constructs, addressing H4--H7. Multiple linear regression models were then estimated to test whether the three mediator constructs jointly predicted each outcome construct, with separate models specified for Learning \& Skill Development (H4), Perceived Value (H5), and Global Citizenship \& Readiness (H6). Predictors were standardized prior to regression to enable comparison of effect magnitudes across constructs. Model diagnostics included variance inflation factors (VIF) \cite{obrien2007} to assess multicollinearity and Shapiro--Wilk tests \cite{shapiro1965} on model residuals to evaluate the normality assumption.
Open-ended responses were analyzed using inductive thematic coding \cite{braun2006}, with codes aligned to the seven constructs. Two additional researchers reviewed coding, resolving discrepancies through discussion. Frequencies were used to identify dominant themes, reported alongside quantitative results.
\section{Results}
\subsection{Reliability Analysis}

All seven constructs exceeded the reliability threshold of
$\alpha \geq .70$, ranging from .757 (CC) to .931 (GCR),
indicating acceptable to excellent internal consistency
(Table~\ref{tab:reliability}). CC1 showed the weakest
item-total correlation across all scales ($r = .361$) and
would marginally improve the CC scale if removed, but was
retained given its content relevance to the construct.

\begin{table}[ht]
\centering
\caption{Reliability of Survey Constructs (N = 36)}
\small
\begin{tabular}{lc}
\hline
\textbf{Construct} & \boldmath$\alpha$ \\
\hline
Collaborative Experience (CE)           & .886 \\
Communication \& Cultural Dynamics (CC) & .757 \\
Logistical Strategies (LS)              & .807 \\
Learning \& Skill Development (LSD)     & .874 \\
Perceived Value (PV)                    & .894 \\
Global Citizenship \& Readiness (GCR)   & .931 \\
Challenges \& Opportunities (CO)        & .870 \\
\hline
\end{tabular}
\label{tab:reliability}
\end{table}
\subsection{Descriptive Statistics}
All construct means exceeded 4.0 on the five-point scale
(Table~\ref{tab:descriptive}), well above the neutral midpoint
of 3.0, indicating consistently positive student ratings across
all dimensions. PV was rated highest ($M = 4.478$) and CC
lowest ($M = 4.067$), though both reflect strong agreement.
Shapiro--Wilk tests confirmed non-normality across all seven
constructs ($p < .05$), consistent with a ceiling effect where
most students rated things positively. Non-parametric tests
were used throughout.

\begin{table}[ht]
\centering
\caption{Descriptive Statistics (N = 36)}
\small
\begin{tabular}{lccc}
\hline
\textbf{Construct} & \textbf{M} & \textbf{SD} & \textbf{Mdn} \\
\hline
CE  & 4.378 & 0.689 & 4.600 \\
CC  & 4.067 & 0.623 & 4.000 \\
LS  & 4.106 & 0.668 & 4.200 \\
LSD & 4.356 & 0.574 & 4.500 \\
PV  & 4.478 & 0.527 & 4.600 \\
GCR & 4.350 & 0.665 & 4.600 \\
CO  & 4.144 & 0.689 & 4.200 \\
\hline
\multicolumn{4}{l}{\footnotesize See Appendix~\ref{tab:survey} for full construct names and item descriptions.} \\
\end{tabular}
\label{tab:descriptive}
\end{table}
\subsection{Group Comparison: U.S. vs. Pakistan}
Mann--Whitney U tests found no significant differences between
U.S. ($n = 21$) and Pakistan ($n = 15$) students on any
construct (all $p > .05$, all effect sizes small). As shown
in Table~\ref{tab:groups}, Pakistan students rated everything
slightly higher across the board, with the largest gap on PV.
Pakistan students also showed notably lower variability
(e.g., $SD_{PV} = 0.183$ vs. $0.628$ for U.S. students),
suggesting a more consistent experience within that cohort.
These results indicate that the positive findings hold across
both institutional contexts.

\begin{table}[ht]
\centering
\caption{Mean Scores by Cohort}
\small
\begin{tabular}{lcc}
\hline
\textbf{Construct} & \textbf{US Mean} & \textbf{PAK Mean} \\
\hline
CE  & 4.295 & 4.493 \\
CC  & 4.010 & 4.147 \\
LS  & 3.981 & 4.280 \\
LSD & 4.305 & 4.427 \\
PV  & 4.314 & 4.707 \\
GCR & 4.219 & 4.533 \\
CO  & 4.038 & 4.293 \\
\hline
\multicolumn{3}{l}{\footnotesize No significant differences (all $p > .05$).} \\
\end{tabular}
\label{tab:groups}
\end{table}
\subsection{Hypothesis Testing: H1--H3}
One-sample Wilcoxon signed-rank tests confirmed that all three
mediator constructs were significantly above neutral, supporting
H1, H2, and H3. As shown in Table~\ref{tab:wilcoxon}, all
effects were large ($r > .82$, all $p < .001$), indicating
that positive ratings were consistent and not driven by a
small number of outliers. All four outcome constructs also
exceeded neutral significantly (all $p < .001$,
$r = .835$--$.872$).

\begin{table}[ht]
\centering
\caption{Wilcoxon Tests Against Neutral Midpoint (N = 36)}
\small
\begin{tabular}{lcccc}
\hline
\textbf{Construct} & \textbf{Mdn} & \textbf{W} & \textbf{Z} & \textbf{r} \\
\hline
CE  (H1) & 4.60 & 620.5 & 5.004 & .834 \\
CC  (H2) & 4.00 & 648.5 & 4.957 & .826 \\
LS  (H3) & 4.20 & 619.0 & 4.979 & .830 \\
LSD      & 4.50 & 666.0 & 5.232 & .872 \\
PV       & 4.60 & 665.0 & 5.216 & .869 \\
GCR      & 4.60 & 657.0 & 5.090 & .848 \\
CO       & 4.20 & 652.0 & 5.012 & .835 \\
\hline
\multicolumn{5}{l}{\footnotesize All $p < .001$. $r$ = effect size (large $\geq$ .50).} \\
\end{tabular}
\label{tab:wilcoxon}
\end{table}
\subsection{Hypothesis Testing: H7}

Students acknowledged that the project involved real challenges
(CO1; $M = 3.917$, above neutral), but consistently rated
personal growth and opportunity items higher
(CO2--CO5; $M = 4.201$). A paired Wilcoxon test confirmed
this difference was significant ($W = 301.5$, $Z = 2.243$,
$p = .012$, $r = .374$, medium effect). The item asking whether the experience was
``overall more rewarding than challenging'' received the
highest mean of all CO items ($M = 4.389$, $Mdn = 5.000$),
with more than half of students giving it the maximum
rating. H7 was supported.

\subsection{Correlational and Regression Analysis: H4--H6}

All 21 Spearman correlations among construct means were
significant ($p \leq .006$). Two relationships stood out:
CC and LS were highly intercorrelated ($\rho = .887$),
reflecting that students who felt communication went well
also felt logistics went well, and PV and GCR moved closely
together ($\rho = .883$). All 12 mediator-outcome pairs were
significant and positive, with 11 of 12 showing large effects
($\rho = .572$--$.760$). The only medium-effect pair was
CE--CO ($\rho = .445$, $p = .006$).

Regression results are in Table~\ref{tab:regression}. All
three models were significant, with the mediators jointly
explaining 54.8\% of variance in LSD, 44.7\% in PV, and
52.1\% in GCR (adjusted $R^2$). No individual predictor reached significance in the LSD
or PV models, likely reflecting both the high intercorrelation between CC and LS ($\rho = .887$) and the modest sample size ($n = 36$), which together limited the model's ability to isolate individual predictor contributions even when the overall model explained a substantial share of the outcome. Shapiro--Wilk testing of model residuals indicated a normality
violation for the LSD model ($W = 0.897$, $p = .003$); the LSD
model coefficients should therefore be interpreted with caution,
though the overall model fit remains valid. In the GCR model, CC was the only significant predictor ($\beta = .669$, $p = .013$), meaning cross-cultural communication specifically drove global readiness beyond what CE and LS contributed.
 \begin{table}[ht]
\centering
\caption{Regression Models: Mediators Predicting Outcomes (N = 36)}
\small
\begin{tabular}{llcc}
\hline
\textbf{Outcome} & \textbf{Predictor} & \boldmath$\beta$ & \textbf{p} \\
\hline
LSD                  & CE & \phantom{-}.355  & .080 \\
($R^2_{adj}$ = .548) & CC & \phantom{-}.049  & .843 \\
                     & LS & \phantom{-}.421  & .055 \\
\hline
PV                   & CE & \phantom{-}.171  & .438 \\
($R^2_{adj}$ = .447) & CC & \phantom{-}.204  & .458 \\
                     & LS & \phantom{-}.378  & .116 \\
\hline
GCR                  & CE & \phantom{-}.148  & .469 \\
($R^2_{adj}$ = .521) & CC & \phantom{-}.669  & $.013^{*}$ \\
                     & LS &          $-.052$ & .811 \\
\hline
\multicolumn{4}{l}{\footnotesize $\beta$ = standardized coefficient.} \\
\multicolumn{4}{l}{\footnotesize $^{*}p < .05$. All models significant at $p < .001$.} \\
\end{tabular}
\label{tab:regression}
\end{table}
\subsection{Qualitative Themes}

Open-ended responses were coded across all seven constructs.
Table~\ref{tab:qualcodes} presents the dominant themes and
their frequencies. A total of 294 coded instances were
identified across 31 responses, with most students
contributing codes to multiple constructs.

Positive Experience was the most frequent CE theme ($n = 22$),
with students broadly describing the collaboration as
enjoyable and worthwhile. Technical Skills Gained was the
highest-frequency code in the entire dataset ($n = 20$),
reflecting widespread self-reported learning gains in
Python, data cleaning, visualization, and machine learning.
Growth Through Struggle dominated the CO construct ($n = 19$),
with students consistently reframing logistical and
interpersonal challenges as productive learning experiences.
Cultural Diversity as Asset ($n = 15$) and Confidence in
Global Work ($n = 12$) were the leading GCR themes,
reinforcing the quantitative finding that cross-cultural
engagement drives global readiness. Time Zone Challenges
($n = 11$) was the most frequently cited specific difficulty,
though it appeared almost exclusively alongside growth-oriented
codes rather than as a standalone negative.

\begin{table*}[ht]
\centering
\caption{Dominant Qualitative Themes by Construct}
\small
\begin{tabular}{llcp{6cm}}
\hline
\textbf{Construct} & \textbf{Theme} & \textbf{n} & \textbf{Description} \\
\hline
CE & Positive Experience              & 22 & Overall satisfaction, willingness to repeat \\
   & Successful Task Despite Barriers & 10 & Project completed despite challenges \\
   & Time Zone Friction               &  4 & Coordination difficulty due to time difference \\
   & Unequal Participation            &  3 & Inactive or dominating partners \\
\hline
CC & Cultural Diversity as Asset      & 15 & Different backgrounds broadened thinking \\
   & Minimal Cultural Difference      & 10 & Culture felt largely invisible to the work \\
   & Time Shaped Communication        &  7 & Time gap dictated when/how people communicated \\
   & Work Focused Interaction         &  3 & Communication stayed task-oriented \\
\hline
LS & Coordination \& Task Management  & 17 & Roles, deadlines, organization done efficiently \\
   & Communication Practices          & 16 & Frequent updates and check-ins via group chat \\
   & Challenges \& Inefficiencies     &  6 & Unclear instructions, tool issues, confusion \\
   & Time Zone \& Scheduling          &  5 & Async delays due to availability \\
\hline
LSD & Technical Skills Gained         & 20 & Improved current skills in data and ML \\
    & Cross-Cultural Collab Skills    & 16 & Adapting to different peers and work styles \\
    & Async \& Prof Communication     & 15 & Clear communication and structured reporting \\
    & Exposure to New Methods         &  7 & New techniques or tools not yet encountered \\
\hline
PV  & Professional Value              & 15 & Tangible career credential, real-world simulation \\
    & Personal Growth                 & 13 & Confidence, open-mindedness, resilience \\
    & Teamwork Transcends Distance    & 10 & Collaboration works regardless of geography \\
    & Academic Value                  & 10 & Deeper subject matter understanding \\
\hline
GCR & Adaptability \& Cultural Intel. & 14 & Situational flexibility and cultural sensitivity \\
    & Confidence in Global Work       & 12 & Preparedness for global environments \\
    & Technical Work as Universal Lang&  5 & Shared tools enabled cross-cultural collaboration \\
\hline
CO  & Growth Through Struggle         & 19 & Resilience and learning through setbacks \\
    & Time Zone Challenges            & 11 & Coordination difficulties and timing conflicts \\
    & Partner Engagement Issues       &  7 & Communication gaps and unresponsive partners \\
\hline
\end{tabular}
\label{tab:qualcodes}
\end{table*}
\section{Discussion}
This study examined student experiences in a short-term VGC embedded in undergraduate computing courses
across two institutions. All seven hypotheses were supported,
and findings were consistent across U.S. and Pakistan cohorts.

\subsection{Students Rated the Experience Positively}

H1 through H3 confirm that students found collaboration
inclusive, communication effective, and logistics manageable,
all well above neutral with large effect sizes. This aligns
with prior work showing that structured virtual exchange,
supported by clear task frameworks and shared digital tools,
enables effective asynchronous collaboration across
institutional and cultural boundaries \cite{vahed2020,
ismailov2021}. The fact that U.S. and Pakistan students rated
everything similarly, despite differences in time zone and
institutional context, suggests the VGC model worked for both
groups rather than just one.

Pakistan students reported slightly higher scores with notably
less variability across all scales. This may reflect stronger
collective motivation among volunteer participants from that
institution, or greater prior familiarity with asynchronous
digital tools.
Qualitative responses reinforce this pattern. Positive
Experience was the most frequent CE code ($n = 22$), and
Successful Task Despite Barriers ($n = 10$) indicates that
even students who encountered friction completed the project
and described it favorably. The small number of Unequal
Participation codes ($n = 3$) is consistent with the lower
tail of the CE score distribution, suggesting that partner
disengagement was an isolated rather than systemic issue.

\subsection{Learning and Value Were Driven Collectively}

The mediators together explained roughly half to nearly
60\% of variance in learning and perceived value, showing
that process quality during collaboration is meaningfully
connected to what students get out of the experience.
Individual predictors did not reach significance in these
models because CC and LS were so closely related in
students' perceptions that the regression could not
separate their individual contributions. Students who
felt communication went well almost always felt logistics
went well too, making the two constructs statistically
inseparable. This suggests that in short-term virtual collaboration,
students may not cognitively distinguish between
communication quality and logistical coordination,
experiencing them instead as a single dimension of
collaboration effectiveness. This pattern extends prior
work on virtual team coordination \cite{davidsonshivers2023} and
points to a practical implication: instructional designs
that strengthen communication in distributed teams may
simultaneously improve logistical outcomes without
targeting them separately.
This interpretation is directly supported by the qualitative
data. Coordination and Task Management ($n = 17$) and
Communication Practices ($n = 16$) were the two dominant LS
codes, and both describe behaviors that are fundamentally
communicative: dividing tasks, sending updates, and
maintaining shared awareness. Students managed logistics
by communicating, making the two constructs functionally
inseparable in practice. Technical Skills Gained ($n = 20$)
and Cross-Cultural Collab Skills ($n = 16$) were the leading
LSD codes, confirming that students could name specific
learning gains and that these gains spanned both technical
and interpersonal dimensions.

\subsection{Communication Drives Global Readiness}

The most notable finding is that CC was the only predictor
that uniquely and significantly explained global citizenship
and readiness ($\beta = .669$, $p = .013$). Students who
engaged more actively with cross-cultural communication
reported stronger confidence in international work and a
deeper sense of global identity, independent of how smooth
their logistics were or how included they felt in the team.
This points to cultural interaction itself as the active
ingredient in building global competence, extending prior
research showing intercultural communication is central to
virtual exchange outcomes \cite{hackett2023, chu2024}.
The qualitative data provides a mechanism for this finding.
Cultural Diversity as Asset ($n = 15$) was the dominant CC
theme, indicating that students who engaged with cultural
difference experienced it as enriching rather than
obstructive. A recurring qualitative theme, Technical Work
as Universal Language ($n = 5$), further suggests that
shared technical tools and methods served as a bridge across
cultural differences, enabling collaboration even where
communication styles diverged. These themes collectively
explain how cross-cultural communication translates into
global readiness: students who engaged with cultural
differences through the shared medium of data science work
developed adaptability and confidence that logistics and
teamwork alone could not produce.

\subsection{Challenges Were Seen as Worth It}

Students did not downplay the difficulties. CO1 was rated
above neutral, meaning students genuinely acknowledged that
challenges were part of the experience. Yet they still came
away feeling the opportunities and personal growth outweighed
those challenges, with most giving the maximum possible
rating to the idea that the project was more rewarding than
difficult. This supports the view that productive struggle
in virtual collaboration contributes to deeper learning
rather than detracting from it.
This is most powerfully illustrated in the qualitative data.
Growth Through Struggle was the highest-frequency CO code
($n = 19$) and appeared across both U.S. and Pakistan
cohorts, with students consistently narrating challenges
as productive rather than purely negative. Time Zone
Challenges ($n = 11$) was the most named specific
difficulty, yet students typically described workarounds
such as shift-based asynchronous workflows and relay
documentation strategies. Partner Engagement Issues
($n = 7$) represented the main interpersonal difficulty.
Together these themes explain why H7 showed a medium
rather than large effect: the challenges were real and
consistently acknowledged, but the reframing toward
growth was equally consistent, producing a meaningful
but moderate advantage for opportunities over challenges.
\section{Limitations}
Several limitations should be noted. Participation was
optional, which may favor students already motivated toward
global collaboration. All measures are self-reported and
reflect a single short activity, limiting causal claims.
The sample size ($n = 36$) reduced regression power and
made group-specific models for each cohort unfeasible.
Multicollinearity between CC and LS limited the
interpretability of individual predictors in the LSD and
PV models, though the overall models remained valid. The
LSD regression also showed a minor residual normality
violation, meaning those results should be read with some
caution even though the model was significant.
\section{Conclusion and Future Work}
This study shows that a short-term, structured virtual
global collaboration can meaningfully support technical
learning, collaborative skill development, and global
readiness in undergraduate computing education. Students
from both the U.S. and Pakistan reported consistently
positive experiences, and the three mediators together
explained roughly 45 to 55\% of outcome variance across
all models (adjusted $R^2$). Cross-cultural communication
emerged as the unique driver of global citizenship and
readiness, pointing to intercultural interaction as the
core mechanism through which this type of collaboration
builds global competence. Qualitative themes reinforced
and extended this finding, with students describing
cultural diversity as an asset and shared technical work
as a bridge across cultural differences, providing a
student-level explanation for why cross-cultural
communication translates into global readiness. These
findings offer a replicable and scalable model for
integrating global learning into computing curricula
without the barriers of traditional study abroad.
Future work should replicate this design with larger
samples to enable group-specific analyses and improve
statistical power. Longitudinal follow-up would help
determine whether gains in global readiness persist
after the collaboration ends. Expanding to additional
institutions and cultural contexts would further test
generalizability. The thematic coding in this study
relied on emergent codes verified by two additional
researchers, but did not employ formal
inter-rater reliability metrics. Future work should
apply structured reliability protocols such as Cohen's
kappa and expand qualitative coding to larger and more
diverse student samples to assess whether the themes
identified here generalize across different institutional
and cultural contexts.
\section*{Acknowledgment}
The authors thank the students from both participating institutions who volunteered for the VGC project. The authors also used an LLM for grammar
correction and improving sentence clarity during manuscript preparation.
\bibliographystyle{IEEEtran}
\bibliography{references}
\appendix

\section{Survey Instrument}

Table~\ref{tab:survey} presents the complete survey instrument.
All Likert items (L) used a five-point scale
(1 = Strongly Disagree, 5 = Strongly Agree).
Open-ended items (OE) were optional.

\begin{table*}[ht]
\centering
\caption{Complete Survey Instrument}
\small
\begin{tabular}{p{2.8cm}llp{9.5cm}}
\hline
\textbf{Construct} & \textbf{Item} & \textbf{Type} & \textbf{Question} \\
\hline

\textbf{Collaborative Experience (CE)}
 & CE1  & L  & I felt included as a full member of the project team. \\
 & CE2  & L  & I was able to contribute meaningfully to discussions and decisions. \\
 & CE3  & L  & My contributions were valued by international teammates. \\
 & CE4  & L  & I was satisfied with the overall collaboration experience. \\
 & CE5  & L  & I felt a sense of teamwork with students from other universities. \\
 & CE6  & OE & Describe your overall experience of working with international peers. \\
\hline

\textbf{Communication \& Cultural Dynamics (CC)}
 & CC1  & L  & Communication within the team was clear and effective. \\
 & CC2  & L  & I felt comfortable expressing my ideas to international peers. \\
 & CC3  & L  & Cultural differences positively influenced our collaboration. \\
 & CC4  & L  & I was able to adapt to different communication styles in the group. \\
 & CC5  & L  & Misunderstandings (if any) were resolved constructively. \\
 & CC6  & OE & What role did cultural differences play in your collaboration? \\
\hline

\textbf{Logistical Strategies (LS)}
 & LS1  & L  & Coordinating across time zones was manageable. \\
 & LS2  & L  & Our team effectively scheduled meetings and task deadlines. \\
 & LS3  & L  & The digital tools and platforms we used supported collaboration well. \\
 & LS4  & L  & Technical or scheduling issues were resolved efficiently. \\
 & LS5  & L  & The team found creative ways to overcome logistical barriers. \\
 & LS6  & OE & What strategies worked best or failed in managing logistics? \\
\hline

\textbf{Learning \& Skill Development (LSD)}
 & LSD1 & L  & I improved my technical and data science knowledge through this project. \\
 & LSD2 & L  & I developed stronger teamwork and collaboration skills. \\
 & LSD3 & L  & I learned new approaches to problem-solving from my peers. \\
 & LSD4 & L  & I improved my communication skills in a virtual environment. \\
 & LSD5 & L  & This project expanded my awareness of how data science is applied globally. \\
 & LSD6 & OE & What skills (technical or non-technical) did you develop through this project? \\
\hline

\textbf{Perceived Value (PV)}
 & PV1  & L  & The global collaboration experience was valuable for my education. \\
 & PV2  & L  & I would participate in a similar project again in the future. \\
 & PV3  & L  & The project provided benefits beyond regular coursework. \\
 & PV4  & L  & I believe this type of experience should be offered in future classes. \\
 & PV5  & L  & The experience positively impacted my academic motivation. \\
 & PV6  & OE & What value did this project bring to you personally, academically, or professionally? \\
\hline

\textbf{Global Citizenship \& Readiness (GCR)}
 & GCR1 & L  & This project increased my confidence in working with diverse, global teams. \\
 & GCR2 & L  & I gained a deeper appreciation for global perspectives in problem-solving. \\
 & GCR3 & L  & I feel more prepared for future international or multicultural work environments. \\
 & GCR4 & L  & The experience enhanced my sense of global citizenship. \\
 & GCR5 & L  & I believe this project contributed to my career readiness in a global workforce. \\
 & GCR6 & OE & How do you think this experience prepared you for working in global contexts? \\
\hline

\textbf{Challenges \& Opportunities (CO)}
 & CO1  & L  & The project involved significant challenges that influenced my experience. \\
 & CO2  & L  & The project provided unique opportunities I would not have had otherwise. \\
 & CO3  & L  & I grew personally from overcoming challenges in the project. \\
 & CO4  & L  & The challenges I faced were balanced by meaningful opportunities. \\
 & CO5  & L  & I consider the overall experience more rewarding than challenging. \\
 & CO6  & OE & What were the biggest challenges you faced? What opportunities did you gain? \\
\hline

\multicolumn{4}{l}{\footnotesize L = Likert scale (1 = Strongly Disagree, 5 = Strongly Agree). OE = Open-ended (optional).} \\
\end{tabular}
\label{tab:survey}
\end{table*}
\end{document}